\documentclass[11pt]{article}
\usepackage{graphicx}

\begin{document}

\def\xslash#1{{\rlap{$#1$}/}}
\def \p {\partial}
\def \dd {\psi_{u\bar dg}}
\def \ddp {\psi_{u\bar dgg}}
\def \pq {\psi_{u\bar d\bar uu}}
\def \jpsi {J/\psi}
\def \psip {\psi^\prime}
\def \to {\rightarrow}
\def\bfsig{\mbox{\boldmath$\sigma$}}
\def\DT{\mbox{\boldmath$\Delta_T $}}
\def\xit{\mbox{\boldmath$\xi_\perp $}}
\def \jpsi {J/\psi}
\def\bfej{\mbox{\boldmath$\varepsilon$}}
\def \t {\tilde}
\def\epn {\varepsilon}
\def \up {\uparrow}
\def \dn {\downarrow}
\def \da {\dagger}
\def \pn3 {\phi_{u\bar d g}}

\def \p4n {\phi_{u\bar d gg}}

\def \bx {\bar x}
\def \by {\bar y}


\begin{center}
{\Large\bf Twist-3 Contributions in Semi-Inclusive DIS with Transversely Polarized Target }
\par\vskip20pt
A.P. Chen$^{1}$, J.P. Ma$^{1,2}$, G.P. Zhang$^{2}$     \\
{\small {\it
$^1$ Institute of Theoretical Physics, Academia Sinica,
P.O. Box 2735,
Beijing 100190, China\\
$^2$ Center for High-Energy Physics, Peking University, Beijing 100871, China  
}} \\
\end{center}
\vskip 1cm
\begin{abstract}
We study semi-inclusive DIS with a transversely polarized target in the approach of collinear factorization. 
The effects related with the transverse polarization are at twist-3. We derive the complete result of twist-3 contributions 
to the relevant hadronic tensor at leading order of $\alpha_s$, and construct correspondingly experimental observables.
Measuring these observables will help to extract the twist-2 transversity distribution, twist-3 distributions and twist-3 fragmentation 
functions of the produced unpolarized hadron. A detailed comparison with the approach of transverse-momentum-dependent factorization 
is made.

\vskip 5mm
\noindent
\end{abstract}
\vskip 1cm
\par 
Experiments of lepton-hadron collisions with large momentum transfers have played an important role in exploring  
the inner structure of hadrons. Typical examples are DIS- and Semi-Inclusive DIS(SIDIS) processes. Based on collinear 
factorizations of QCD, the differential cross-sections of DIS and SIDIS at the leading power 
are predicted with parton distributions of the initial hadron and fragmentation function of the produced hadron. These distributions and fragmentation functions are defined as matrix elmemets  
of QCD twist-2 operators. In this letter we study the contributions involving twist-3 operators in SIDIS.      
\par 
We will assume that the polarization of the produced hadron is not observed. The twist-3 contributions 
in SIDIS appear only in the case that the initial hadron is transversely polarized. 
The contributions  contain not only twist-3 matrix elements of the initial hadron introduced in \cite{EFTE,QiuSt}, 
but also the twist-2 transversity distribution introduced in \cite{JaJi}, combined with chirality-odd twist-3 fragmentation functions. 
Therefore, they contain rich information about the inner structure 
of hadrons. Experimentally, the twist-3 contributions can be measured through asymmetries caused by the transverse spin.  Hence, those twist-3 distributions and fragmentation functions can be extracted from the asymmetries. The relevant studies  
in experiments planned in the future can be found in \cite{JLab1,EIC1,EXP1} and references therein.   

\par   
SIDIS with transversely polarized target has been studied in \cite{EKT,EKT2,KaKo}. In these works one assumes that the initial 
lepton is unpolarized and the hadron in the final state has large transverse momentum. The obtained Single transverse-Spin Asymmetry(SSA) starts at order of $\alpha_s$. We will derive the complete result of the twist-3 hadronic tensor of SIDIS at order of $\alpha_s^0$. With the complete results we construct spin-dependent observables at $\alpha_s^0$.
Through measuring these observables one can extract relevant parton distributions and fragmentation functions. It is 
interesting to note that the obtained twist-3 hadronic tensor at tree-level can be expressed completely with the parton distributions and fragmentation functions defined with two-parton correlations.    
\par 
With the employed approach 
of collinear factorization one can only derive the twist-3 hadronic tensor as a distribution tensor of the transverse 
momentum. From the tensor one can only obtain physical predictions in which the transverse momentum is integrated. At tree-level the produced hadron has  a small transverse-momentum at order of $\Lambda_{QCD}$. In this kinematical 
region one can employ the approach of Transverse-Momentum-Dependent(TMD) factorization studied in \cite{CS,CSS,JMY}. The complete angular 
distribution of SIDIS at tree-level has been derived with the approach in \cite{PMRT,DBPM,DBJPM} and in \cite{NSIDIS}. We will discuss 
the difference between the two approaches in detail after giving our results.     

\par

We consider the SIDIS process: 
\begin{equation} 
   e(k,\lambda_e) + h(P,s) \to e(k') + h'(P_h) + X
\label{Eq1} 
\end{equation} 
where the initial hadron is of spin-1/2 with the spin vector $s$. The initial electron can be 
polarized with the helicity $\lambda_e$. The polarization of the hadron in the final state is not observed. At leading order of QED, there is an exchange of one virtual photon with the momentum $q=k-k'$ between the electron and the initial hadron. 
The relevant hadronic tensor as:
\begin{equation} 
W^{\mu\nu} = \sum_X \int \frac {d^4 x}{(2\pi)^4} e^{iq\cdot x} \langle P,s\vert J^\mu (x) \vert P_h, X\rangle 
     \langle X, P_h\vert J^\nu (0) \vert P, s\rangle. 
\end{equation} 
The standard variables for SIDIS are:
\begin{equation} 
x_B = \frac{Q^2}{2 P\cdot q},\ \ \  y=\frac{ P\cdot q}{P\cdot k}, \ \ \ z_h=\frac{P\cdot P_h}{P\cdot q}. 
\end{equation} 
We will neglect the masses of hadrons and leptons.    

\par
       
It is convenient to use the  light-cone coordinate system, in which a
vector $a^\mu$ is expressed as $a^\mu = (a^+, a^-, \vec a_\perp) =
((a^0+a^3)/\sqrt{2}, (a^0-a^3)/\sqrt{2}, a^1, a^2)$. Two light-cone vectors are introduced as 
$l^\mu =(1,0,0,0)$ and $n^\mu=(0,1,0,0)$. With these two vectors one can define two transverse tensors: 
$g_\perp^{\mu\nu} = g^{\mu\nu} - n^\mu l^\nu - n^\nu l^\mu$ and $\epsilon_\perp^{\mu\nu} =\epsilon^{\alpha\beta\mu\nu}l_\alpha n_\beta$. With these notations we introduce the relevant parton distributions and fragmentation functions.  In this work we will use Feynman gauge. 

\par 
Assuming that the initial hadron moves in the $z$-direction with the momentum $P^\mu=(P^+,0,0,0)$ and it is transversely 
polarized with $s^\mu=(0,0,s_\perp^1,s_\perp^2)$, the transversity distribution is defined as\cite{JaJi}: 
\begin{equation} 
  h_1 (x)  s_\perp^\mu = \int \frac{ d\lambda }{4\pi} e^{-ix \lambda P^+} \langle P,s_\perp \vert \bar \psi(\lambda n) {\mathcal L}^\dagger_n(\lambda n)  \gamma^+ \gamma^\mu_\perp\gamma_5 
  {\mathcal L}_n(0) \psi(0) \vert P,s_\perp\rangle  
\end{equation} 
where ${\mathcal L}_n(\xi)$ is the gauge link starting from $\xi$ to $\infty$ in space-time. The transversity distribution is of twist-2. At twist-3 one can define the three twist-3 distributions from two-parton correlations:    
\begin{eqnarray} 
  q_T (x) s_\perp^\mu  &=&  P^+ \int \frac{d\lambda}{ 4\pi } e^{- i x\lambda  P^+} \langle P,s_\perp \vert \bar \psi(\lambda n) {\mathcal L}^\dagger_n(\lambda n)  \gamma_\perp^\mu \gamma_5  
  {\mathcal L}_n(0) \psi(0) \vert P,s_\perp\rangle, 
\nonumber\\  
  -i q_\partial (x) s_\perp^\mu &=&    \int \frac{d\lambda}{ 4\pi } e^{- i x\lambda  P^+} \langle P,s_\perp \vert \bar \psi(\lambda n) {\mathcal L}^\dagger_n(\lambda n)  \gamma^+ \gamma_5 \partial_\perp^{\mu}   
  \left ( {\mathcal L}_n  \psi \right ) (0) \vert P,s_\perp\rangle,
\nonumber\\  
  -i q_\partial' (x) \tilde s_\perp^\mu &=&    \int \frac{d\lambda}{ 4\pi } e^{- i x\lambda  P^+} \langle P,s_\perp \vert \bar \psi(\lambda n) {\mathcal L}^\dagger_n(\lambda n)  \gamma^+ \partial_\perp^{\mu}   
  \left ( {\mathcal L}_n  \psi \right ) (0) \vert P,s_\perp\rangle.    
\label{tw31}  
\end{eqnarray}
One may replace in the first line of Eq.(\ref{tw31}) $\gamma_5$ with $I$ to define another twist-3 distribution. But 
one can show  that it is zero\cite{MaZh2}. The three distributions are real. From  three-parton correlations one can define two twist-3 distributions:   
\begin{eqnarray} 
T_F (x_1,x_2)  \tilde s_{\perp}^\mu  
   &=&    g_s  \int \frac{d\lambda_1 d\lambda_2}{4\pi}
   e^{ -i\lambda_2 (x_2-x_1) P^+ -i \lambda_1 x_1 P^+ }
   \langle P, \vec s_\perp \vert
           \bar\psi (\lambda_1n ) \gamma^+ G^{+\mu}(\lambda_2n) \psi(0) \vert P,\vec s_\perp \rangle ,  
\nonumber\\
T_{\Delta} (x_1,x_2) s_\perp^\mu 
   & =&  - i g_s  \int \frac{d\lambda_1 d\lambda_2}{4\pi}
   e^{ -i\lambda_2 (x_2-x_1) P^+ -i \lambda_1 x_1 P^+ }
 \langle P, \vec s_\perp \vert
           \bar\psi (\lambda_1n )  \gamma^+ \gamma_5   G^{+\mu}(\lambda_2n)  \psi(0) \vert P,\vec s_\perp \rangle, 
\label{tw3}
\end{eqnarray}
where we have suppressed the gauge links for short notations and $\tilde s^\mu =\epsilon_\perp^{\mu\nu} s_{\perp\nu}$. Corresponding to the two distributions in Eq.(\ref{tw3})  
one can define additionally two twist-3 distributions by replacing the field strength tensor $g_s G^{+\mu}(x)$ 
with $P^+ D^\mu_\perp(x)$, where $D^\mu(x)$ is given by $D^\mu (x)=\partial_\mu + ig_s G^\mu (x)$. These two functions 
will not appear in our calculation. In fact they can be expressed with the distributions given 
in Eq.(\ref{tw31},\ref{tw3}) as shown in \cite{EKT}.  Among the introduced five twist-3 distributions one can show:   
\begin{equation} 
\frac{1}{2\pi}\int d x_1  P\frac{ 1 }{x_1-x_2}  \biggr [ T_F (x_1,x_2)  - T_\Delta (x_1,x_2) \biggr ] = -x_2 q_T (x_2) + q_\partial (x_2), \quad
   T_F (x,x) = - 2 q'_\partial (x).   
\label{RL1} 
\end{equation}   
where $P$ stands for the principle-value prescription. The first relation has been derived in \cite{JiG2}.  The second relation  
is obtained by examining the relation between $T_F$ and that obtained from $T_F$ by the mentioned replacement. It should be emphasized that 
the second relation is for SIDIS. We note here that the distribution $q'_\partial (x)$ is defined with the gauge links in Eq.(\ref{tw31}) pointing to the future to factorize the effects of final-state interactions in SIDIS. In Drell-Yan processes there are no final-state interactions. But there are initial-state interactions. Hence, 
The distribution $q'_\partial (x)$ in Drell-Yan processes is defined with gauge links pointing to the past. 
With the symmetries of time-reversal and parity one can show that there is a sign-difference between the two distributions. 
For Drell-Yan processes, the $-$-sign in the relation should be replaced with $+$. We notice here that the two distributions 
with different gauge links have been studied in \cite{BMPT}, where it has been shown that the difference between the two distributions is proportional to $T_F(x,x)$. The twist-3 distribution $q'_\partial (x)$ can also be defined as a transverse-momentum-moment of Sivers function. Such moments are in general related to parton distributions at high twists, as discussed 
in \cite{GRVT}.

\par
To define fragmentation functions, we assume  that the produced hadron moves in the $-z$-direction with the momentum $P^\mu =(0,P^-,0,0)$. From two-parton correlations we define:  
\begin{eqnarray} 
  && z P^- \int\frac{d \xi}{2\pi} e^{- i \xi P^-/z} \sum_X   \biggr [  
    \langle  0\vert {\mathcal L}^\dagger _l (0)\psi(0)  \vert P X \rangle_i \langle  X P \vert \bar \psi(\xi l){\mathcal L}_l (\xi l) \vert 0 \rangle_j \biggr ]
 \nonumber\\   
   && =  \biggr (\gamma\cdot P \hat d(z)  +  \hat e (z) + \frac{i}{2} \sigma_{\alpha\beta} \gamma_5 \epsilon_\perp^{\alpha\beta} \hat e_I(z)   \biggr )_{ij} 
      +\cdots ,
\label{2PFF}         
\end{eqnarray}  
where $ij$ stand for Dirac indices and color indices. ${\mathcal L}_l (\xi)$
is the gauge link along the direction $l^\mu$ starting from $-\infty$ to $\xi$ in space-time. 
$\hat d(z)$ is the standard twist-2 fragmentation function\cite{PDFFF}.  $\hat e$ and $\hat e_I$ ar twist-3 fragmentation functions introduced in  \cite{JiFF}.  Besides these two twist-3 fragmentation functions, there are another two twist-3 fragmentation functions defined as:
 \begin{eqnarray} 
 \hat E_F (z_1,z_2) &=& - \frac{z_2 g_s}{4 N_c}  \int \frac{d\lambda_1 d\lambda_2}{(2\pi)^2} e^{-\lambda_1 P^-/z_1 -i\lambda_2P^-(1/z_2 -1/z_1) }
  \epsilon_{\perp\mu\nu} 
\nonumber\\  
   && \sum_X {\rm Tr} \langle 0\vert \gamma^- \gamma^\nu \gamma_5 \psi(0) \vert h X\rangle_i \langle h X \vert \bar \psi (\lambda_1 l ) G^{-\mu}(\lambda_2 l) 
   \vert 0 \rangle ,
\nonumber\\
  \hat e_\partial (z) &=& -i\frac{z}{4 N_c}   \int \frac{d\lambda}{2\pi} e^{-\lambda P^+/z }
   \sum_X  {\rm Tr} \langle 0\vert \gamma^- \gamma^\nu\gamma_5   {\mathcal L}_l ^\dagger (0) \psi_i(0) \vert h X\rangle \partial_\perp^\mu    \langle h X \vert \bar \psi (\lambda l ){\mathcal L}_l (\lambda l)   \vert 0 \rangle \epsilon_{\perp\mu\nu}.   
\label{3PT3}       
\end{eqnarray}
Similarly one can define an additional fragmentation function $\hat E_D$ by replacing 
$g_s G^{-\mu}(\lambda_2l)$ with $P^- D_\perp^\mu (\lambda_2 l)$. But this function 
is completely determined by $\hat E_F$ and $\hat e_\partial$\cite{MPF}.  All introduced 
twist-3 fragmentation functions are chirality-odd. The functions
$\hat e$, $\hat e_I$ and $\hat e_\partial$ are real, while $\hat E_F$ is complex in general. 
It is shown in \cite{MPF} that there are relations among these four twist-3 fragmentation functions. In our notations they are: 
\begin{equation}
 z_2^2 \int \frac{d z_1}{z_1} P\frac{1}{ z_2-z_1} {\rm Im } \hat E_F (z_1,z_2) = z_2 \hat e_\partial (z_2) - \hat e_I(z_2), 
 \quad \hat e(z_2) = z_2^2 \int \frac{d z_1}{z_1} P\frac{1}{z_2-z_1} {\rm Re }\hat E_F(z_1,z_2). 
\label{RL2}
\end{equation}  
In \cite{MeMe} it is shown that $\hat E_F(z,z)=0$. This implies that there will be no soft-gluon-pole contributions represented by $\hat E_F(z,z)$. 
\par 
For deriving the twist-3 contribution to $W^{\mu\nu}$ for the process in Eq.(\ref{Eq1}), it is convenient to take the frame,
in which the initial hadron moves in the $z$-direction with $P^\mu=(P^+,0,0,0)$ and the final hadron moves 
in the $-z$-direction with $P_h^\mu =(0, P_h^-,0,0)$.  Then the virtual photon has the momentum $q^\mu =(q^+,q^-,q_\perp^1,q_\perp^2)$.  We call this frame as 
${\mathcal C}_0$- frame. 
The obtained result is covariant. It can be conveniently transformed into the frame called 
${\mathcal C}_1$-frame, in which the virtual photon moves in the $-z$-direction and 
the initial hadron moves in the $z$-direction. The produced hadron in ${\mathcal C}_1$-frame
can then have  nonzero transverse momentum.  

\par 

\par
\begin{figure}[hbt]
\begin{center}
\includegraphics[width=14cm]{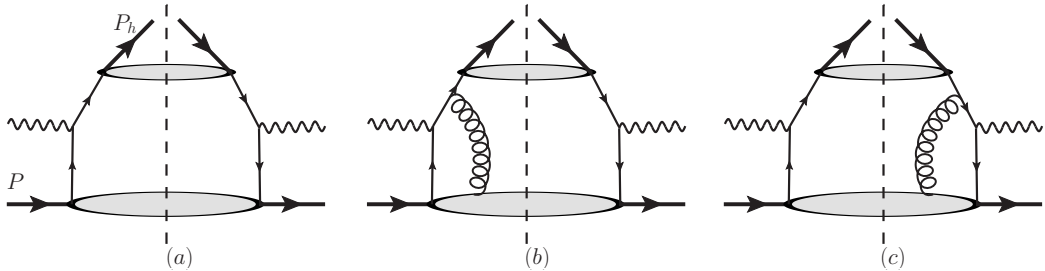}
\end{center}
\caption{Diagrams for contributions in SIDIS. } 
\end{figure}
\par

The twist-3 $W^{\mu\nu}$ can be devided into two parts: One consists of contributions 
with nonperturbative quantities defined with chirality-odd operators, another one 
consists of contributions with nonperturbative quantities defined with chirality-even operators.  The chirality-even part can only contain the twist-2 fragmentation function and twist-3 parton distributions.  At tree-level, it receives contributions from diagrams given in Fig.1. 
It is rather standard to calculate contributions at different twists from Fig.1 by collinear expansion. We take Fig.1a as an example to illustrate this. 
\par 
Within the power accuracy considered here, one can already neglect the $+$- and the transverse 
components of momenta carried by the parton lines entering into the upper bubble in Fig.1. 
One can also neglect the $-$-components of momenta carried by the parton lines entering into the lower bubble. Projecting out the twist-2 part related to the final hadron, the contribution from Fig.1a can be written as: 
\begin{eqnarray}
W^{\mu\nu} \biggr\vert_{1a} &=& \int d k_B^- d k_A^3 \biggr [  \delta^4(q +k_A-k_B) 
\frac{1}{z} \hat d(z) \biggr ( \gamma^\mu \gamma^+  \gamma^\nu \biggr )_{ij} \biggr ] 
 \cdot \int \frac{d^3 \xi }{ (2\pi)^3} e^{ik_A\cdot\xi } \langle h(P) \vert \bar q_i (0) q_j (\xi ) \vert 
   h(P) \rangle, 
\nonumber\\
      && \xi^\mu = (0,\xi^-,\vec \xi_\perp),\quad k_A^\mu = (k_A^+, 0,\vec k_{A\perp}),\quad  k_B^\mu=(0,k_B^-,0,0) =(0,P_h^-/z,0,0), 
\label{F1a}           
\end{eqnarray}     
where $ij$ stand for Dirac- and color indices. $k_A$ is the momentum 
carried by the quark line leaving the lower bubble in the left of Fig.1a, $k_B$  is the momentum 
carried by the quark line entering into the upper bubble. If we neglect $k_{A\perp}$ 
in $[\cdots ]$ in Eq.(\ref{F1a}), we obtain the twist-2 contribution. One needs to expand 
the $[\cdots ]$ in Eq.(\ref{F1a}) in $k_{A\perp}$ and to make corresponding projections of the quark density matrix to obtain the twist-3 contribution. After the expansion and projections we have: 
\begin{eqnarray} 
W^{\mu\nu} \biggr\vert_{1a}  &=&  -i \delta^2(q_\perp) \frac{1}{z_h  } \hat d(z_h) \epsilon^{\mu\nu\alpha\beta}  n_\alpha    \int \frac{d\xi^- }{ 2\pi } e^{i \xi^- x P^+} \langle h(P) \vert \bar q(0) \gamma_{\perp\beta} \gamma_5  q (\xi^-n  ) \vert 
   h(P) \rangle
\nonumber\\
    && -  \frac{\partial}{\partial q_\perp^\rho} \delta^2(q_\perp) \frac{1}{z_h  } \hat d(z_h) \epsilon_\perp^{\mu\nu} \int \frac{d\xi^- }{ 2\pi } e^{i \xi^- x P^+} \langle h (P) \vert \bar q(0) \gamma^+ \gamma_5 \partial_\perp^\rho  q (\xi^-n  ) \vert 
   h(P) \rangle
\nonumber\\
    && -  \frac{\partial}{\partial q_\perp^\rho} \delta^2(q_\perp) \frac{1}{z_h  } d(z_h) g_\perp^{\mu\nu}  i \int \frac{d\xi^- }{ 2\pi } e^{i \xi^- x P^+} \langle h \vert \bar q(0) \gamma^+  \partial_\perp^\rho  q (\xi^-n  ) \vert 
   h \rangle   
    + \cdots , 
\label{F1A3}    
\end{eqnarray} 
where $\cdots$ stand for contributions at twist-2 or beyond twist-3.  
The three correlation functions of quark fields in Eq.(\ref{F1A3}) 
look like the three distributions $q_T$, $q_\partial$ and $q'_\partial$ defined in Eq.(\ref{tw31}) without 
the gauge links. If one considers the contributions from Fig.1b and 1c and those with exchanges of more than one gluon,
one can realizes 
that parts of contributions from exchanges of gluons can be summed into gauge links. Adding these parts to the contributions 
in the above, the results are simply obtained by replacing the three correlation functions in Eq.(\ref{F1A3})  with $q_T$, $q_\partial$ and  $q'_\partial$,  respectively.

\par 
It is straightforward to calculate the contributions from Fig.1b and Fig.1c. Parts of the contributions will be added to the contribution of Fig.1a as discussed in the above. 
The remaining contributions can be easily found as: 
 \begin{eqnarray}
W^{\mu\nu}\biggr\vert_{1b+ 1c} &=& -\frac{1}{z_h} \hat d(z_h) \delta^2 (q_\perp) \frac{i }{\pi P\cdot q}  \biggr ( P^\mu \tilde s_\perp^\nu - P^\nu \tilde s_\perp^\mu \biggr )   \int  d x_1 P\frac{1 }{x_1-x_B}  \biggr [ T_F (x_1,x_B)  - T_\Delta (x_1,x_B) \biggr ]
\nonumber\\ 
  && + \frac{1}{z_h} d(z_h) \delta^2 (q_\perp) \frac{1 }{ k_B^-}  \biggr ( l^\mu \tilde s_\perp^\nu + l^\nu \tilde s_\perp^\mu \biggr ) T_F(x_B,x_B) 
+ \cdots,
\label{F1BC3}    
\end{eqnarray} 
where $\cdots$ stand for contributions beyond twist-3. The symmetric part is obtained by the absorptive part of the quark propagator between the photon- and gluon vertex in Fig.1b and Fig.1c. This gives the so-called soft-gluon pole contribution.    We note here that the results in Eq.(\ref{F1A3},\ref{F1BC3})  
can be simplified with the relation in Eq.(\ref{RL1}). Adding every contributions we have the total chirality-even part of $W^{\mu\nu}$:   
\begin{eqnarray}
W^{\mu\nu}\biggr\vert_{even} &=&  \frac{2}{z_h} \hat d(z_h) \delta^2 (q_\perp) \frac{i }{P\cdot q }  \biggr ( P^\mu \tilde s_\perp^\nu - P^\nu \tilde s_\perp^\mu \biggr )  \biggr ( x_B q_T (x_B) - q_\partial (x_B) \biggr ) 
\nonumber\\
   && -i \delta^2(q_\perp) \frac{2}{z_h P\cdot P_h }\hat d(z_h) \epsilon^{\mu\nu\alpha\beta}  P_{h\alpha}  s_{\perp\beta} q_T (x_B)  +  i \frac{2}{z_h  }\hat d(z_h) \epsilon_\perp^{\mu\nu}  q_\partial (x_B)  s_\perp^\rho \frac{\partial}{\partial q_\perp^\rho} \delta^2(q_\perp) 
\nonumber \\
   && + \frac{1}{z} \hat d(z) T_F (x_B,x_B) \biggr [ \delta^2 (q_\perp) \frac{1}{P\cdot q } ( P^\mu \tilde s_\perp^\nu 
   + P^\nu \tilde s_\perp^\mu ) + g_\perp^{\mu\nu} \tilde s_\perp^\rho \frac{\partial}{\partial q_\perp^\rho} \delta^2(q_\perp)   \biggr ] .  
\label{CEVEN}   
\end{eqnarray} 
The symmetric part was first derived in \cite{MaZh2}.
Before we turn to the chirality-odd part, we discuss the $U(1)$-gauge invariance of our result. 
From the invariance one always has $q_\mu W^{\mu\nu}=0$. But our $W^{\mu\nu}$ is singular in $q_\perp$. It can only be taken as a distribution tensor. Then the $U(1)$-gauge invariance implies that for any test function  ${\mathcal T}(q_\perp)$ one has: 
\begin{equation} 
  \int d^2 q_\perp {\mathcal T}(q_\perp) q_\mu W^{\mu\nu}  =0.  
\label{U1}   
\end{equation}       
It is easy to check that our result in Eq.(\ref{CEVEN}) is $U(1)$-gauge invariant.  
\par
\begin{figure}[hbt]
\begin{center}
\includegraphics[width=14cm]{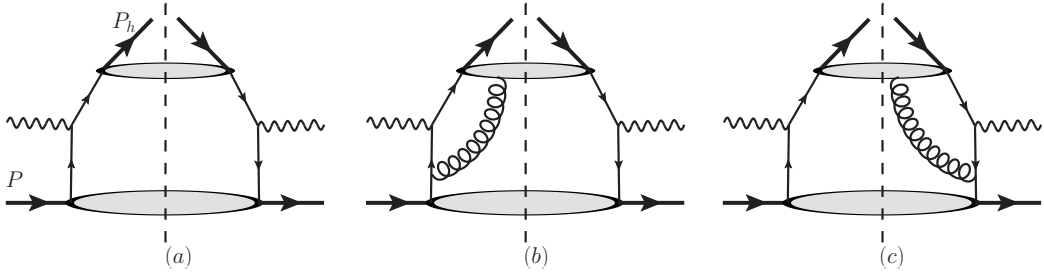}
\end{center}
\caption{Diagrams for contributions in SIDIS. } 
\end{figure}
\par 

\par
The chirality-odd contribution involves the transversity distribution and twist-3 fragmentation 
functions. It receives the contributions from diagrams given in Fig.2.  The calculation of Fig.2
is similar to that of Fig.1.  Here we omit the details of derivation and give the results directly:
\begin{eqnarray} 
W^{\mu\nu}\biggr\vert_{2a} &=& 
  \biggr [  i\epsilon^{\mu\nu\alpha\beta} P_\alpha s_{\perp \beta} \hat e(z_h) - (P^\mu\tilde s_\perp^\nu + P^\nu \tilde s_\perp^\mu ) \hat e_I (z_h) \biggr ]\delta^2 (q_\perp)  \frac{2}{z_h P\cdot P_h} h_1(x_B)  
\nonumber\\  
  &&  +  \biggr [g_\perp^{\mu\nu} \tilde s_\perp^\rho - g_\perp^{\mu\rho}\tilde s_\perp^\nu 
       - g_\perp^{\nu\rho}\tilde s_\perp^\mu \biggr ] h_1 (x_B)  \hat e_\partial (z_h)  \frac{1}{z_h} \frac{\partial}{\partial q_\perp^\rho} \delta^2 (q_\perp),
\nonumber\\
W^{\mu\nu}\biggr\vert_{2b+2c} &=& -\frac{2}{ x_B P\cdot P_h} \biggr (P_h^\nu \tilde s^\mu_\perp + P_h^\mu\tilde s^\nu_\perp 
  \biggr )  \delta^2 ( q_\perp)  h_1(x_B) \int \frac{dz_1}{z_1} P\frac{1}{z_1-z_h} {\rm Im } \hat E_F(z_1,z_h)
\nonumber\\
    && +  \frac{2 i}{ x_B P\cdot P_h} \biggr (P_h^\nu \tilde s^\mu_\perp - P_h^\mu\tilde s^\nu_\perp \biggr )  \delta^2 ( q_\perp)  h_1(x_B) \int \frac{dz_1}{z_1} P\frac{1}{z_1-z_h} {\rm Re }\hat E_F(z_1,z_h).
\label{FIG2}      
\end{eqnarray} 
In the above only twist-3 contributions are given explicitly. Parts of contributions from Fig.2b and 2c are added into the contributions to Fig.2a as we have done for Fig.1.
The contribution from Fig.2b and 2c can be simplified with the relation in Eq.(\ref{RL2}).  
With the relation we obtain the chiral-odd contribution: 
\begin{eqnarray} 
 W^{\mu\nu} \biggr\vert_{odd} &=&   \frac{2 i}{z_h P\cdot P_h} h_1 (x_B) \hat e(z_h) \delta^2 (q_\perp) \biggr [ 
   \epsilon^{\mu\nu\alpha\beta} P_\alpha s_{\perp \beta} - \frac{1}{x_B z_h} 
   \biggr (P_h^\nu \tilde s^\mu_\perp - P_h^\mu\tilde s^\nu_\perp \biggr ) \biggr ] 
\nonumber\\
  && - \frac{2} {z_h P\cdot P_h} \delta^2 (q_\perp) h_1(x_B) \biggr [  (P^\mu\tilde s_\perp^\nu + P^\nu \tilde s_\perp^\mu ) \hat e_I (z_h)  - \frac{1}{ x_B z_h } \biggr (P_h^\nu \tilde s^\mu_\perp + P_h^\mu\tilde s^\nu_\perp 
  \biggr ) \biggr ( z_h \hat e_\partial (z_h)  
\nonumber\\  
  && - \hat e_I (z_h) \biggr ) \biggr ] + \biggr [g_\perp^{\mu\nu} \tilde s_\perp^\rho -g_\perp^{\mu\rho}\tilde s_\perp^\nu 
       -g_\perp^{\nu\rho}\tilde s_\perp^\mu \biggr ] h_1 (x_B) \hat e_\partial (z_h)  \frac{1}{z_h} \frac{\partial}{\partial q_\perp^\rho} \delta^2 (q_\perp).  
\label{CODD}          
\end{eqnarray} 
It is easy to find with Eq.(\ref{U1}) that  the above result is $U(1)$-gauge invariant. 
The total twist-3 contribution of $W^{\mu\nu}$ is then the sum of the chirality-even part 
in Eq.(\ref{CEVEN}) and the chirality-odd part in Eq.(\ref{CODD}).  

\par 
To study how to construct experimental observables it is convenient to express our 
$W^{\mu\nu}$ in the introduced ${\mathcal C}_1$-frame where the produced hadron has nonzero transverse momentum. The transverse momentum $P_{h\perp}$ in the  ${\mathcal C}_1$-frame is related to the transverse momentum $q_\perp$ in the  ${\mathcal C}_0$-frame
as: 
\begin{equation} 
   q_\perp^\mu \biggr\vert_{{\mathcal C}_0} = -\frac {1}{z_h} P_{h\perp}^\mu\biggr\vert_{{\mathcal C}_1}. 
\end{equation}    
It should be noted that the two tensors $g_\perp^{\mu\nu}$ 
and $\epsilon_\perp^{\mu\nu}$  are covariant. But they are defined 
differently in different frames. In the following we will use the same notations for momenta and 
spin without confusion. The two tensors in the ${\mathcal C}_1$-frame are defined as:
\begin{equation}
g_\perp^{\mu\nu} = g^{\mu\nu} -\frac{1}{P\cdot \bar P} \left ( P^\mu \bar P^\nu 
 + P^\nu \bar P^\mu\right ),\quad \epsilon_\perp^{\mu\nu} =  \frac{1}{P\cdot \bar P} \epsilon^{\alpha\beta\mu\nu} P_\alpha \bar P_\beta , \quad \bar P = x_B P +q. 
\end{equation}  
With these notations our twist-3 contribution of $W^{\mu\nu}$ in the  ${\mathcal C}_1$-frame
is given by: 
\begin{eqnarray}
W^{\mu\nu}&=&  \frac{2} {x_B P\cdot q}\delta^2 (P_{h\perp})  \biggr [  - i \epsilon^{\mu\nu\alpha\beta} q_\alpha s_{\perp\beta}    
     \biggr (  x_B z_h  q_T (x_B) \hat d(z_h)  +h_1 (x_B) \hat e(z_h)  \biggr )
\nonumber\\
     && +\biggr ( (2 x_B P+q)^\mu \tilde s_\perp^\nu +  (2 x_B P+q)^\nu\tilde s_\perp^\mu \biggr )    h_1(x_B)  \biggr (  z_h \hat  e_\partial (z_h) -\hat e_I (z_h) \biggr ) \biggr ]   
\nonumber\\
   && + z_h^2  \left (  \frac{\partial}{\partial P_{h\perp}^\rho} \delta^2(P_{h\perp} ) \right ) \biggr [ - 2 i q_\partial (x_B) \hat d(z_h)    \epsilon_\perp^{\mu\nu} s_\perp^\rho  
     - T_F(x_B,x_B) \hat d(z_h)   g_\perp^{\mu\nu} \tilde s_\perp^\rho 
\nonumber\\              
 && - \biggr ( g_\perp^{\mu\nu} \tilde s_\perp^\rho -g_\perp^{\mu\rho}\tilde s_\perp^\nu 
       -g_\perp^{\nu\rho}\tilde s_\perp^\mu \biggr ) h_1 (x_B) \hat e_\partial (z_h)  \biggr ] ,     
\label{TOTA}   
\end{eqnarray} 
This expression is explicitly $U(1)$-gauge invariant because in the ${\mathcal C}_1$-frame
$q^\mu$ is given by $q^\mu = (q^+,q^-,0,0)$.  This result is our main result. 
The obtained $W^{\mu\nu}$ in Eq.(\ref{TOTA}) 
is in fact a tensor distribution of $P_{h\perp}$, physical predictions can only be obtained 
if $P_{h\perp}$ is integrated out.
We notice that in our main result all nonperturbative quantities are defined with two-parton 
correlations with the relation between  $T_F(x_B,x_B)$ and $q_\partial' (x_B)$ in Eq.(\ref{RL1}). 
However, this is only true at tree-level. Beyond 
tree-level it is not the case as discussed in detail in \cite{JiG2}.  
                         
\par 
We consider the experimental situation in which the initial hadron is transversely polarized 
with the spin vector $s_\perp^\mu$. This vector is transverse to the lepton beam direction. 
In fact, this vector is not exactly the transverse spin vector in the ${\mathcal C}_1$-frame. But, in 
the kinematical region of large $Q^2$, the two vectors are approximately the same\cite{Diehl}.  
We will neglect the difference between the two spin vectors.  The incoming- and outgoing lepton span the so-called lepton plane. In the ${\mathcal C}_1$-frame 
the azimuthal angle between the spin vector and the lepton plane is denoted $\phi_s$. Similarly, one defines the azimuthal angle $\phi_h$ for the produced hadron.
The azimuthal angle of the outgoing lepton around the lepton beam with respect to the spin vector is denoted $\psi$.
In the kinematical region of SIDIS, one has $\psi\approx \phi_s$\cite{Diehl}. With this specification the differential cross-section is given by\cite{NSIDIS,Diehl}:
\begin{equation} 
   \frac{ d\sigma}{d x_B d y d z_h d\psi d^2  P_{h\perp} } = \frac{\alpha^2 y}{4 z_h Q^4} L_{\mu\nu} W^{\mu\nu}. 
\end{equation}
As discussed before, from our result we cannot predict the differential cross-section as a function of $\phi_h$ and $P_{h\perp}^2$. 
The predictions can only be made by integrating out $P_{h\perp}$.  
Using the result in Eq.(\ref{TOTA}) and integrating $P_{h\perp}$ out, we obtain the twist-3 contribution to the differential cross-section:
\begin{eqnarray}
   \frac{ d\sigma  }{d x_B d y d z_h d\psi }     &=& \frac{4 \alpha^2 }{z_h Q^3 }\vert s_\perp \vert\sqrt{1-y} \biggr [ -\lambda_e x_B  \biggr (  x_B z_h q_T (x_B)\hat d(z_h) +h_1 (x_B) \hat e(z_h)  \biggr ) \cos\psi
\nonumber\\    
     && +\frac{2-y}{y} h_1(x_B)  \biggr (  z_h \hat e_\partial (z_h) -\hat e_I (z_h) \biggr ) \sin\psi \biggr ]. 
\label{dsig}             
\end{eqnarray} 
In this angular distribution the terms in Eq.(\ref{TOTA}) with the derivative of $\delta^2(P_{h\perp})$ do not contribute. 
To extract information of these terms, one can construct weighted angular distributions defined as:
\begin{eqnarray}
   \frac{ d\sigma \langle {\mathcal F} \rangle }{d x_B d y d z_h d\psi} = \frac{\alpha^2 y}{4 z_h Q^4}  \int d^2 P_{h\perp} L_{\mu\nu} W^{\mu\nu}  
       {\mathcal F} (P_h, k',s_\perp )         
\end{eqnarray} 
with ${\mathcal F}$ as the weight function. For ${\mathcal F}=1$ one obtains the differential cross-section given in Eq.(\ref{dsig}). One can obtain the following weighted angular distribution for these derivative terms:
\begin{eqnarray} 
\frac{ d\sigma \langle P_{h\perp}\cdot k'_\perp  \rangle }{d x_B d y d z_h d\psi} &=&  \frac{\alpha^2 z_h }{2 Q y^2}   
\vert s_\perp \vert\sqrt{1-y}  \biggr [ -2 \lambda_e y (2-y)   q_\partial (x_B) \hat d(z_h) \cos\psi 
\nonumber\\
    && -  \biggr (  (1+(1-y)^2)   T_F(x_B,x_B) \hat d(z_h) 
  - 2 (1-y) h_1 (x_B) \hat e_\partial (z_h) \biggr ) \sin\psi   \biggr ].
\label{dsigF}    
\end{eqnarray} 
One can construct more spin-dependent observables by integrating over the azimuthal angle with different weight functions.
Our $W^{\mu\nu}$ in Eq.(\ref{TOTA}) have five tensor structures. Correspondingly one can have five observables. We can 
obtain the five weighted differential cross-sections: 
\begin{eqnarray} 
\frac{ d\sigma \langle P_{h\perp}\cdot\tilde s_\perp \rangle }{d x_B d y d z_h} &=& \frac{\pi \alpha^2}{Q^2} \frac{1 +(1-y)^2}{y} 
  \vert s_\perp \vert ^2 z_h  T_F(x_B,x_B)\hat d(z_h) , 
\nonumber\\
\frac{ d\sigma  \langle P_{h\perp}\cdot s_\perp \rangle }{d x_B d y d z_h} &=& - \lambda_e \frac{2 \pi \alpha^2 (2-y) }{Q^2} 
  \vert s_\perp \vert ^2 z_h  q_\partial (x_B) \hat d(z_h), 
\nonumber\\
\frac{ d\sigma \langle P_{h\perp}\cdot k_\perp' k_\perp'\cdot \tilde s_\perp \rangle }{d x_B d y d z_h} &=& 
    \pi\alpha^2 \vert s_\perp \vert ^2z_h \frac{(1-y)^2}{y^3} \biggr [ h_1(x_B) \hat e_\partial (z_h) - 
    \frac{1+(1-y)^2 }{2(1-y)} \hat d(z_h) T_F(x_B,x_B)   \biggr ] ,
\nonumber\\
\frac{ d\sigma \langle k_\perp'\cdot s_\perp \rangle }{d x_B d y d z_h} &=& \lambda_e\frac{4\pi \alpha^2}{z_h Q^2} 
  \frac{1-y}{y} 
  \vert s_\perp \vert ^2 \biggr (  x_B z_h  q_T (x_B) \hat d(z_h) +h_1 (x_B) \hat e(z_h)  \biggr ) , 
\nonumber\\
\frac{ d\sigma  \langle k_\perp' \cdot \tilde s_\perp \rangle }{d x_B d y d z_h} &=& \frac{4\pi \alpha^2  }{z_h Q^2}\frac{1-y}{y^2} (2-y)  
  \vert s_\perp \vert ^2 h_1(x_B)  \biggr (  z_h \hat  e_\partial (z_h) -\hat e_I (z_h) \biggr ) .    
\label{WSSA}             
\end{eqnarray}   
There are corrections to our hadronic tensor given in Eq.(\ref{TOTA}). They are from higher orders of $\alpha_s$ and 
from power-corrections suppressed by $1/Q$ or $1/Q^2$.  Therefore, our results of observables in Eq.(\ref{dsig}, \ref{dsigF}, 
\ref{WSSA}) are subjects of these corrections. Parts of one-loop corrections to the weighted differential cross section 
with $P_{h\perp}\cdot \tilde s_\perp$ in Eq.(\ref{WSSA}) have been given in \cite{KVX,KVX2}. 

\par 
Now we are in position to discuss the difference in our case between TMD- and collinear factorization. In general, TMD factorization 
can be used in the kinematical region of $P_{h\perp}\ll Q$. If one has $Q \gg P_{h\perp} \gg \Lambda_{QCD}$, 
one can show that the TMD factorization and the collinear 
factorization are equivalent in this kinematical region\cite{JQVY1,JQVY2}.  At first look, one can 
use these two factorization approaches to calculate observables like those in Eq.(\ref{WSSA}), in which $P_{h\perp}$ is integrated out. However,  this is not trivial in fact.    
We will take the contribution 
containing Sivers-function as an example to show this. 
\par 
The relevant contribution in TMD factorization is\cite{NSIDIS}:
\begin{eqnarray} 
\frac{d\sigma}{d x_B d y d z_h d\psi d^2 P_{h\perp} } &=& \frac{ \alpha^2 (1 + (1-y)^2) }{x_B y Q^2}  \vert s_\perp\vert 
 \sin(\phi_h-\phi_s)   F_{UT,T}^{\sin(\phi_h-\phi_s) } (x_B, z_h, P_{h\perp}) 
\nonumber\\ 
    && \cdot \biggr [ 1 + {\mathcal O} (P_{h\perp}^2/Q^2) \biggr ] +\cdots, 
\nonumber\\
   F_{UT,T}^{\sin(\phi_h-\phi_s) }(x_B, z_h, P_{h\perp})  &=&\frac{ x_B}{ \vert P_{h\perp}\vert} \int d^2 k_{A\perp} d^2 k_{B\perp} 
          P_{h\perp}\cdot  k_{A\perp} f_{1T}^\perp (x_B,k_{A\perp})  D_1 (z_h,k_{B\perp}) 
\nonumber\\         
        && \cdot  \delta^2 (k_{A\perp}-k_{B\perp}-P_{h\perp}/z_h),       
\label{SIVERS}           
\end{eqnarray}  
where $\cdots$ denote irrelevant contributions. The contribution given explicitly is relevant to the weighted differential 
cross section with $P_{h\perp}\cdot\tilde s_\perp$ in Eq.(\ref{WSSA}). We take the same notations here as those in \cite{NSIDIS}. $f_{1T}^\perp (x,k_\perp)$ 
is the Sivers function of the initial hadron and $D_1(z,k_\perp)$ is the TMD fragmentation function. Their definitions can be found 
in \cite{NSIDIS}.  Formally, one can derive the relations:
\begin{equation}
  \hat d(z) = z^2 \int d^2 k_\perp D_1 (z,k_\perp), \quad  \int d^2 k_\perp \vert k_\perp\vert^2 f_{1T}^\perp (x,k_\perp) = 
     -T_F(x,x).
\label{TMCL}       
\end{equation}
The second equation is derived in \cite{BMPT}.      
With TMD factorization one is able to predict the distribution in $\phi_h$. But, the contribution given 
in Eq.(\ref{SIVERS}) can only be used in the kinematical region for $P_{h\perp}\ll Q$. It has power corrections. 
If we neglect the power-corrections, from Eq.(\ref{SIVERS}) we have for the weighted differential cross section:  
\begin{eqnarray} 
\frac{ d\sigma \langle P_{h\perp}\cdot\tilde s_\perp \rangle }{d x_B d y d z_h}  
      &=& - \frac{\pi \alpha^2}{Q^2} \frac{1 +(1-y)^2}{ y} 
  \vert s_\perp \vert ^2  z_h \int d^2 P_{h\perp}d^2 k_{A\perp} d^2 k_{B\perp}  \vert k_{A\perp}\vert^2 f_{1T}^\perp (x_B,k_{A\perp}) 
       D_1 (z_h,k_{B\perp}) 
\nonumber\\         
        && \cdot  \delta^2 (k_{A\perp}-k_{B\perp}-P_{h\perp}/z_h).         
\end{eqnarray}    
If we use the $\delta$-function to perform the integration of $P_{h\perp}$, and then take the integration of $k_{A\perp}$ 
and that of $k_{B\perp}$ as two independent integrals, one can obtain the same result as that in Eq.({\ref{WSSA}) 
by using the relation in Eq.(\ref{TMCL}). However, in principle one can not derive it in this way 
with TMD factorization. In fact the integration of $k_{A\perp}$ and that of $k_{B\perp}$ are not independent. They are 
correlated. Kinematically $P_{h\perp}$ is always finite. It 
can not be infinitely large. Therefore $k_{A\perp}-k_{B\perp}$ is always finite. 
 Since $P_{h\perp}$ here with TMD factorization is constrained at the order of $\Lambda_{QCD}$ with $\Lambda_{QCD} \ll Q$, one always 
has the constraint $k_{A\perp}-k_{B\perp} \sim \Lambda_{QCD}$.  The only way to derive the same result is to assume that one can neglect $k_{A\perp}$ and 
$k_{B\perp}$ in the $\delta$-function. But with this assumption it implies that one actually uses collinear factorization 
at the beginning. It is interesting to note that at tree-level neglecting $k_{A\perp}$ and 
$k_{B\perp}$ in the $\delta$-function is equivalent to relaxing the constraint $k_{A\perp}-k_{B\perp}\sim \Lambda_{QCD}$. Keeping these in mind, our results of observables can also be derived from TMD factorization.  It is also worth to point out here that the formally derived relations in Eq.(\ref{TMCL}) are not exactly 
correct, as discussed in \cite{KQVY}. The reason is that in the integrations over transverse momenta there will be U.V. divergences and 
an U.V. subtraction needs to be implemented. This can be shown with the explicit calculation 
of $f_{1T}^\perp$ and $T_F(x,x)$ at the leading order of $\alpha_s$ with a multi-parton state in \cite{MS3}.  
\par 
Before summarizing our work, we point out that from the derivation of our hadronic tensor in Eq.(\ref{TOTA}) 
one can realize that the virtual corrections to the terms with the derivative of $\delta^2(P_{h\perp})$ are determined 
completely by the quark form factor with certain subtractions of collinear divergences. This fact has been first noticed 
in the study of Drell-Yan processes in \cite{MaZh2}.     
\par
To summarize: We have derived the twist-3 part of the hadronic tensor in SIDIS at tree-level. This part depends on the transverse spin 
of the initial hadron. At tree-level, the obtained twist-3 part is completely expressed with nonperturbative quantities 
defined with two-parton correlations. Spin-dependent observables are constructed based on the obtained hadron tensor. A comparison of collinear factorization  with TMD factorization is given in the studied case. 
Measurements of the various spin-dependent observables in SIDIS will be helpful to extract information about 
the transversity distribution at twist-2 and twist-3 parton distributions and fragmentation functions.

\par\vskip20pt
\noindent
{\bf Acknowledgments}
\par
The work of J.P. Ma is supported by National Nature
Science Foundation of P.R. China(No.11275244). The partial support from the CAS center for excellence in particle 
physics(CCEPP) is acknowledged.

\par

\end{document}